\documentclass[conference]{IEEEtran}
\IEEEoverridecommandlockouts
\usepackage{cite}
\usepackage{hyperref}
\usepackage{siunitx}
\usepackage{eurosym}
\DeclareSIUnit{\sieuro}{\mbox{\euro}}
\DeclareSIUnit{\siyear}{y}
\usepackage{orcidlink}
\usepackage{amsmath,amssymb,amsfonts}
\usepackage{algorithmic}
\usepackage{graphicx}
\usepackage{amsmath}
\usepackage{comment}
\usepackage{amssymb}
\usepackage{tikz}
\usepackage{tikz-cd}
\usepackage{pgfplots}
\usepackage{multicol}
\usepackage{csvsimple}
\usetikzlibrary{positioning}
\usepgfplotslibrary{groupplots}
\usepackage{pgfplotstable}
\usepgfplotslibrary{groupplots}
\usepackage{xstring}
\pgfplotsset{compat=1.18}
\usetikzlibrary{positioning,calc,matrix,shapes,shadings}
\usetikzlibrary{spy}
\usetikzlibrary{calc}
\usepgfplotslibrary{statistics, groupplots}
\usetikzlibrary{shapes.geometric}
\usepackage{textcomp}
\usepackage{xcolor}
\definecolor{dkred}{rgb}{0.8,0,0}
\definecolor{blue}{rgb}{0,0,1}

\def\BibTeX{{\rm B\kern-.05em{\sc i\kern-.025em b}\kern-.08em
    T\kern-.1667em\lower.7ex\hbox{E}\kern-.125emX}}
\begin{document}

\title{Simplification Ad Absurdum? Revisiting Gas Flow Modeling for Integrated Energy System Planning
\thanks{This work is part of the project iKlimEt (FO999910627), which has received funding in the framework of ”Energieforschung”, a research and technology program of the Klima- und Energiefonds.}
}

\author{\IEEEauthorblockN{Thomas Klatzer, Yannick Werner, Sonja Wogrin}
\IEEEauthorblockA{\textit{Institute of Electricity Economics and Energy Innovation} \\
\textit{Research Center for Energy Economics and Energy Analytics} \\
\textit{Graz University of Technology}\\
Graz, Austria \\
\{thomas.klatzer, yannick.werner, wogrin\}@tugraz.at}
}

\maketitle

\begin{abstract}
This paper analyzes the implications of simplified pipeline gas flow models for integrated energy system planning. A case study of an integrated power-hydrogen expansion planning problem shows that simplifying pressure-flow relationships and gas dynamics can lead to expansion plans that incur substantial regret when evaluated under a more realistic dynamic gas flow model -- due to suboptimal system expansion, operation, and non-supplied hydrogen.
Numerical experiments show that planning under the highly simplified transport and transport-linepack models -- commonly used in expansion studies -- can result in regret exceeding several thousand percent and yield expansion plans that lack robustness across demand levels. Planning under steady-state conditions partially mitigates these effects, but still leaves significant cost-reduction potential untapped compared to dynamic planning due to neglected linepack flexibility. Developing efficient solution algorithms for the dynamic model is a promising direction for future research.
\end{abstract}


\section{Introduction}
\label{sec:Intro}
The transition towards decarbonized energy systems increasingly emphasizes integrated planning of the power and gas sectors~\cite{EUEnergySystemIntegration2020} to support large-scale production of renewable hydrogen~\cite{EURFNBO2023} and transmission via pipelines~\cite{EHB2023}. Planning of the underlying power–gas infrastructure is commonly supported by integrated energy system optimization models (ESOMs). Within ESOMs, modeling pipeline gas flows is particularly challenging, as gas transport is governed by nonlinear and non-convex pressure–flow relationships and slow dynamics that enable short-term storage via linepack~\cite{Raheli2025}.

Against this background, pipeline expansion planning -- traditionally addressed by the gas systems community -- has received increasing attention in the power systems community. Due to differing objectives, however, the problem is treated differently across the two fields.
The gas systems community typically focuses on optimal network design and does not explicitly consider coupled decisions on gas production or storage, e.g.,~\cite{Andre2013,Baufume2013,Borraz-Sanchez2016}. Here, steady-state gas flow models are state-of-the-art for expansion planning, while dynamic models are used for operational control~\cite{Koch2015,Lenz2021}.
In contrast, the power systems community focuses on integrated expansion of generation, storage, and transmission across power and gas sectors. Gas flow modeling ranges from dynamic~\cite{Zeng2017}, to steady-state~\cite{Borraz-Sanchez2016a}, to highly simplified transport models that approximate linepack~\cite{Bodal2024,He2021}, or omit it entirely~\cite{Zhao2018,Neumann2023}. Due to its computational efficiency, the latter is commonly used in ESOMs such as PyPSA~\cite{Brown2018} and LEGO~\cite{Wogrin2022}.

Systematic comparisons of modeling approaches across different levels of gas flow representation in expansion planning remain scarce. One notable exception is~\cite{Reuss2019}, which analyzes pipeline expansion and hydrogen reconversion in a star- and tree-shaped network under a discrete arc-sizing model, comparing steady-state and transport formulations. However, the study relies on exogenous assumptions for power generation and hydrogen production capacities and does not consider dynamic gas flow representations.
Overall, a comprehensive comparison of gas flow modeling choices for expansion planning and a systematic evaluation of commonly used simplifications are missing. In this work, we address this gap by analyzing the implications of considering (or omitting) technical gas flow characteristics on the integrated expansion of renewable-powered electrolyzers and hydrogen transmission pipelines. Specifically, our contributions are twofold:

(1) We compile and harmonize a range of pipeline gas flow models within a single modeling framework, enabling straightforward comparison across different levels of technical detail -- from a dynamic model that captures nonlinear pressure–flow relationships and slow gas dynamics (linepack) to a highly simplified transport model.
(2)	We conduct in-depth analyses of expansion outcomes for a stylized yet realistic power-hydrogen system and illustrate the implications of planning decision obtained from simplified gas flow models across demand levels when validated under dynamic conditions.

The remainder of the paper is organized as follows: 
Section~\ref{sec:Method} details the hydrogen expansion planning problem, subject to different gas flow modeling choices. In Section~\ref{sec:Experiments}, the numerical experiments and results are presented. Finally, Section~\ref{sec:Conclusions} concludes the paper.

\section{Methodology}
\label{sec:Method}
This section presents the methodology: Subsection~\ref{sec:Model} introduces the hydrogen expansion planning problem, Subsection~\ref{sec:Prerequisites} the gas flow prerequisites, and Subsections~\ref{sec:DynamicModel}--\ref{sec:LinearTransportModelLinepack} the gas flow modeling choices for pipeline expansion planning.

\subsection{Hydrogen Expansion Planning Problem}
\label{sec:Model}
We formulate the hydrogen expansion planning (HEP) problem assuming the perspective of a central planner whose objective is to minimize the total annualized investment and operating cost for hydrogen production and transmission via pipelines over one year.

Let $\mathcal{T}$, $\mathcal{I}$, $\mathcal{E}$, $\mathcal{A}$, $\mathcal{D}$, and $\mathcal{C}$ denote the sets of time steps, nodes, electrolyzers, pipelines, pipeline diameters, and compressors, indexed by $t$, $i$, $e$, $a$, $d$, and $c$, respectively, where the cardinality of $\mathcal{T}$ is denoted by $|\mathcal{T}|$. Pipelines and compressors connect node pairs $i,j\in\mathcal{I}$, such that $\mathcal{A},\mathcal{C}\subseteq\mathcal{I}\times\mathcal{I}$. Electrolyzers are assigned to nodes using the mapping set $\mathcal{M}\subseteq\mathcal{E}\times\mathcal{I}$, where $(e,i)\in\mathcal{M}$ if electrolyzer $e$ is located at node $i$.
Let $x_e$ denote the number of installed electrolyzers, $p^\mathrm{H}_{t,e}$ their hydrogen production (\qty{}{\tonne/\hour}), and $p^\mathrm{Ns}_{t,i}$ the non-supplied hydrogen (\qty{}{\tonne/\hour}). Pipeline expansion with diameter $d$ is represented by $x_{a,d}$.
Pipeline inflows and outflows (\qty{}{\tonne/\hour}) are denoted by $f^\mathrm{In}_{t,a}$ and $f^\mathrm{Out}_{t,a}$, respectively, with the corresponding average gas flow given by $f_{t,a}$. Compressor inflows and outflows (\qty{}{\tonne/\hour}) are denoted by $f^\mathrm{In}_{t,c}$ and $f^\mathrm{Out}_{t,c}$. Electrolyzer capacity (\qty{}{\MW}) is given by the time-variant parameter $\overline{P}^\mathrm{E}_{t,e}$, reflecting the availability of renewable electricity, $\eta_e$ denotes the hydrogen-per-electricity conversion efficiency, and $\overline{X}_e$ the maximum number of electrolyzers. Hydrogen demand (\qty{}{\tonne/\hour}) is denoted by $D^\mathrm{H}_{t,i}$.
Annualized investment costs are denoted by $C^\mathrm{Inv}_e$ for electrolzyers (\qty{}{\sieuro}) and $C^\mathrm{Inv}_{a,d}$ for pipeline expansion (\qty{}{\sieuro}) with diameter $d$. Operating costs comprise electricity costs (\qty{}{\sieuro/\MW\hour}) for hydrogen production $C^\mathrm{E}_{t,e}$ and a penalty (\qty{}{\sieuro/\tonne}) for non-supplied hydrogen $C^\mathrm{Ns}$. Finally, $\Delta$ denotes the time resolution (\qty{}{\hour}).
The HEP problem is formulated as follows:
\begin{subequations}
\label{eqn:HEP}
\begin{align}
    \min & 
      \sum_{e\in\mathcal{E}} C^\mathrm{Inv}_e x_e 
    + \sum_{a\in\mathcal{A}} \sum_{d\in\mathcal{D}} C^{\mathrm{Inv}}_{a,d} x_{a,d} \nonumber\\
    &  + \frac{8760}{\Delta\cdot\mathcal{|T|}} \sum_{t\in\mathcal{T}} \Delta \Big(
         \sum_{e\in\mathcal{E}} C^\mathrm{E}_{t,e} p^{\mathrm{H}}_{t,e}/\eta_e
       + \sum_{i\in\mathcal{I}} C^\mathrm{Ns}  p^{\mathrm{Ns}}_{t,i}
    \Big)
    \label{eqn:Obj} \\
\text{s.t.}    &   \sum_{a(j,i)\in\mathcal{A}} f^{\mathrm{In}}_{t,a} - \sum_{a(i,j)\in\mathcal{A}} f^{\mathrm{Out}}_{t,a} 
    +   \sum_{c(j,i)\in\mathcal{C}} f^{\mathrm{In}}_{t,c} \nonumber \\
    & - \sum_{c(i,j)\in\mathcal{C}} f^{\mathrm{Out}}_{t,c} 
    + \sum_{e:(e,i)\in\mathcal{M}} p^{\mathrm{H}}_{t,e} + p^{\mathrm{Ns}}_{t,i}
    = D^{\mathrm{H}}_{t,i} 
    \quad \forall t,i
    \label{eqn:BalanceH2Simple} \\
    & 0\leq p^{\mathrm{H}}_{t,e} \leq \eta_e \overline{P}^{\mathrm{E}}_{t,e} x_e
    \quad \forall t,e
    \label{eqn:ProdH2} \\
    & 0\leq x_e \leq\overline{X}_e \quad \forall e
    \label{eqn:ELInvBound} \\
    &x_{a,d} \in \{0,1\}
    \quad \forall a,d
    \label{eqn:Loop_Def_Exp} \\
    &\sum_{d\in\mathcal{D}} x_{a,d} = 1
    \quad \forall a
    \label{eqn:Loop_Exp} \\
    &  \text{gas flow modeling choice}~\eqref{eqn:Dynamic_MINLP} \vee \eqref{eqn:Steady_MINLP} \vee \eqref{eqn:TP_Disc_Flow} \vee \eqref{eqn:TP_Disc_Linepack}
    \label{eqn:GasFlowModelingChoice}
\end{align} 
\end{subequations}

For the HEP problem~\eqref{eqn:HEP}, the objective function is given by~\eqref{eqn:Obj}. Constraint~\eqref{eqn:BalanceH2Simple} enforces the hydrogen balance, while~\eqref{eqn:ProdH2} limits hydrogen production from electrolyzers to their expanded capacity. The number of installed electrolyzers is constrained by~\eqref{eqn:ELInvBound}. Pipeline expansion is modeled as a binary decision variable~\eqref{eqn:Loop_Def_Exp}, with at most one diameter selected per pipeline~\eqref{eqn:Loop_Exp}.
Gas transmission~\eqref{eqn:GasFlowModelingChoice} in the HEP problem can be modeled at different levels of physical detail and dynamics, which are introduced in the following Subsections, starting with the prerequisites for gas flow modeling.

\subsection{Gas Flow Modeling Prerequisites}
\label{sec:Prerequisites}
Following standard assumptions in the literature, e.g.,~\cite{Koch2015,Lenz2021}, gas flow in ESOMs is modeled assuming horizontal isothermal pipelines using a system of nonlinear partial differential equations (PDEs) describing conservation of momentum~\eqref{eqn:Momentum} and mass~\eqref{eqn:Mass},
\begin{subequations}
\label{eqn:Model}
\begin{align}
    \frac{\partial p}{\partial x} + \frac{c^2 \lambda }{2DA^2} \frac{f |f|}{p} = 0,
    \label{eqn:Momentum} \\
     \frac{\partial p}{\partial t} \frac{A}{c^2} + \frac{\partial f}{\partial x} = 0,
    \label{eqn:Mass}
\end{align} 
\end{subequations}
where~$x$ and~$t$ denote space and time, and the unknowns are the gas flow~$f$ (\qty{}{\kg/\second}) and gas pressure~$p$ (\qty{}{\bar}). The parameters~$D$ (\qty{}{\meter}) and~$A$ (\qty{}{\meter\squared}) represent the pipeline diameter and cylindrical cross section, while~$c$ deontes the speed of sound, which we define as~$c=\qty{1320}{\meter/\second}$\! for hydrogen. The friction factor~$\lambda$ captures friction-induced pressure losses in pipeline gas flow and is commonly approximated using the Nikuradse formula. Finally, $|\cdot|$ denotes the absolute value.

For use in ESOMs, the PDEs are discretized in space and time using the pipeline length $L$ (\qty{}{\meter}) and the time resolution $\Delta$ as the spatial and temporal step sizes, respectively.

\subsection{Dynamic Model}
\label{sec:DynamicModel}
The resulting discretized PDEs define the dynamic model.
Let $\pi_{t,i}$ denote the gas pressure (\qty{}{\bar}) at node $i\in\mathcal{I}$. For compactness, the parameters $R_{a,d} = \frac{16 c^2 L_{a} \lambda_d}{\pi^2 D^5_{a,d}}$ and $H_{a,d} = \frac{8c^2}{\pi L_a D_{a,d}^2}$ condense pipeline characteristics, where $L_a$, $D_{a,d}$, and $\lambda_d$ denote the pipeline length (\qty{}{\meter}), diameter (\qty{}{\meter}), and corresponding friction factor, respectively, while $c$ and $\pi$ are constants. Moreover, $\underline{\Pi}_i$ and $\overline{\Pi}_i$ represent the minimum and maximum nodal gas pressure (\qty{}{\bar}), respectively, and $\Gamma_c$ is the compression ratio of compressor $c$.
The \textbf{dynamic model} is formulated as:
\begin{subequations}
\label{eqn:Dynamic_MINLP}
\begin{align}
    &\big(\pi^2_{t,i} - \pi^2_{t,j}\big)  = \sum_{d\in\mathcal{D}} R_{a,d} x_{a,d}\ f_{t,a}|f_{t,a}|
    \quad \forall t,a
    \label{eqn:Dynamic_MINLP_Flow} \\
    &f_{t,a} = \frac{1}{2} \Big(f^{\mathrm{In}}_{t,a} + f^{\mathrm{Out}}_{t,a}\Big)
    \quad \forall t,a
    \label{eqn:Dynamic_MINLP_InflowOutflow} \\
    &\pi_{t,i} + \pi_{t,j} = \pi_{t-1,i} + \pi_{t-1,j}     \nonumber \\
    &+ \Delta \sum_{d\in\mathcal{D}} \! H_{a,d} x_{a,d} \Big(f^{\mathrm{In}}_{t,a} - f^{\mathrm{Out}}_{t,a}\Big) 
    \quad \forall~t\in\mathcal{T}\backslash\{1\},a
    \label{eqn:Dynamic_MINLP_Linepack} \\
    &\pi_{t,i} + \pi_{t,j} = \pi_{T,i} + \pi_{T,j}     \nonumber \\
    &+ \Delta \sum_{d\in\mathcal{D}} \! H_{a,d} x_{a,d} \Big(f^{\mathrm{In}}_{t,a} - f^{\mathrm{Out}}_{t,a}\Big) 
    \quad \forall~t\in\{1\},a
    \label{eqn:Dynamic_MINLP_Last_Linepack} \\
    &\pi_{t,j} = \Gamma_c \pi_{t,i}
    \quad \forall t,c
    \label{eqn:Dynamic_MINLP_Comp} \\
    &\underline{\Pi}_{i} \leq \pi_{t,i} \leq \overline{\Pi}_{i}
    \quad \forall t,i
    \label{eqn:Dynamic_MINLP_Press_Bounds}
\end{align}
\end{subequations}

In~\eqref{eqn:Dynamic_MINLP}, constraint~\eqref{eqn:Dynamic_MINLP_Flow} represents the momentum equation governing bidirectional gas flow, defined as the average of inflow and outflow in~\eqref{eqn:Dynamic_MINLP_InflowOutflow}. Constraint~\eqref{eqn:Dynamic_MINLP_Linepack} enforces mass conservation across consecutive time steps, thereby capturing a pipeline’s short-term storage capability via linepack, while~\eqref{eqn:Dynamic_MINLP_Last_Linepack} cyclically links the final and first time step. Constraint~\eqref{eqn:Dynamic_MINLP_Comp} describes pressure increases from compressors, and~\eqref{eqn:Dynamic_MINLP_Press_Bounds} imposes lower and upper bounds on nodal pressures. The dynamic model constitutes a mixed-integer nonlinear program (MINLP) with time-linking constraints and is the most computationally demanding among the models considered.

\subsection{Steady-state Model}
\label{sec:SteadyStateModel}
The dynamic model~\eqref{eqn:Dynamic_MINLP} can be simplified by assuming stationary gas flow, i.e., inflow equals outflow. Consequently, the mass balance constraints~\eqref{eqn:Dynamic_MINLP_Linepack} and~\eqref{eqn:Dynamic_MINLP_Last_Linepack} are omitted. In this setting, pressure variables are expressed as squared nodal gas pressure (\qty{}{\bar\squared}), denoted by~$\pi^{\mathrm{Sqr}}_{t,i}$, with corresponding bounds $\underline{\Pi}^{\mathrm{Sqr}}_i$ and $\overline{\Pi}^{\mathrm{Sqr}}_i$. The \textbf{steady-state model} is formulated as:
\begin{subequations}
\label{eqn:Steady_MINLP}
\begin{align}
    & \big(\pi^{\mathrm{Sqr}}_{t,i} - \pi^{\mathrm{Sqr}}_{t,j}\big) = \sum_{d\in\mathcal{D}} R_{a,d} x_{a,d}\ f_{t,a}|f_{t,a}|
    \quad \forall t,a
    \label{eqn:Steady_MINLP_Flow} \\
    &f^{\mathrm{In}}_{t,a} = f^{\mathrm{Out}}_{t,a} = f_{t,a}
    \quad \forall t,a
    \label{eqn:Steady_MINLP_InflowOutflow} \\
    & \pi^{\mathrm{Sqr}}_{t,j} = \Gamma^2_c \pi^{\mathrm{Sqr}}_{t,i}
    \quad \forall t,c
    \label{eqn:Steady_MINLP_Comp} \\
    & \underline{\Pi}^{\mathrm{Sqr}}_{i} \leq \pi^{\mathrm{Sqr}}_{t,i} \leq \overline{\Pi}^{\mathrm{Sqr}}_{i}
    \quad \forall t,i
    \label{eqn:Steady_MINLP_Press_Bounds}
\end{align}
\end{subequations}

In~\eqref{eqn:Steady_MINLP}, constraint~\eqref{eqn:Steady_MINLP_Flow} represents the momentum equation governing bidirectional gas flow, with inflow equal to outflow~\eqref{eqn:Steady_MINLP_InflowOutflow}. Constraint~\eqref{eqn:Steady_MINLP_Comp} defines the pressure increase from compressors, while~\eqref{eqn:Steady_MINLP_Press_Bounds} imposes lower and upper bounds on nodal pressures. The steady-state model constitutes a MINLP.

\subsection{Transport Model}
\label{sec:DiscreteTransportModel}
The steady-state model~\eqref{eqn:Steady_MINLP} can be approximated as a transport model by omitting the explicit pressure–flow relationship and bounding gas flows by $\overline{F}_{a,d}= \sqrt{\frac{\overline{\Pi}^2_a}{R_{a,d}}}$, corresponding to the maximum flow implied by the maximum pressure $\overline{\Pi}_a$ of pipeline $a$.
The \textbf{transport model} is formulated as:
\begin{subequations}
\label{eqn:TP_Disc}
\begin{align}
    &- \sum_{d\in\mathcal{D}}\overline{F}_{a,d} x_{a,d} \leq f^{\mathrm{In}}_{t,a},f^{\mathrm{Out}}_{t,a} \leq \sum_{d\in\mathcal{D}}\overline{F}_{a,d} x_{a,d}
    \quad \forall t,a
    \label{eqn:TP_Disc_Flow} \\
    &f^{\mathrm{In}}_{t,a} = f^{\mathrm{Out}}_{t,a}
    \quad \forall t,a
    \label{eqn:TP_Disc_InflowOutflow}
\end{align}
\end{subequations}

In~\eqref{eqn:TP_Disc}, constraints~\eqref{eqn:TP_Disc_Flow} limit bidirectional inflow and outflow, while~\eqref{eqn:TP_Disc_InflowOutflow} enforces their equality. The transport problem constitutes a mixed-integer linear program (MILP).

\subsection{Transport Model with Linepack Approximation}
\label{sec:LinearTransportModelLinepack}
The transport model~\eqref{eqn:TP_Disc} can be extended to approximate linepack storage. Let $m_{t,a}$ denote the gas stored in pipeline $a$ at time $t$. The minimum and maximum linepack are given by $\underline{M}_{a,d} = \frac{\pi L_a \underline{\Pi}_a D_{a,d}^{2}}{4c^2}$ and $\overline{M}_{a,d} = \frac{\pi L_a \overline{\Pi}_a D_{a,d}^{2}}{4c^2}$, where $\underline{\Pi}_a$ and $\overline{\Pi}_a$ denote the minimum and maximum pipeline pressures (\qty{}{\bar}). The \textbf{transport-linepack model} is formulated as:
\begin{subequations}
\label{eqn:TP_Disc_Linepack}
\begin{align}
    &- \! \! \sum_{d\in\mathcal{D}} \overline{F}_{a,d} x_{a,d} \leq f^{\mathrm{In}}_{t,a},f^{\mathrm{Out}}_{t,a} \leq \! \! \sum_{d\in\mathcal{D}}\overline{F}_{a,d} x_{a,d}
    \quad \forall t,a
    \label{eqn:TP_Disc_Flow_Linepack} \\
    &m_{t,a} = m_{t-1,a} + \Delta \big(f^{\mathrm{In}}_{t,a} - f^{\mathrm{Out}}_{t,a}\big)
    \quad \forall~t\in\mathcal{T}\backslash\{1\},a
    \label{eqn:TP_Disc_Def_Linepack} \\
    & m_{t,a} = m_{T,a} + \Delta \big(f^{\mathrm{In}}_{t,a} - f^{\mathrm{Out}}_{t,a}\big)
    \! \quad t\in\{1\},~\forall~a
    \label{eqn:TP_Disc_Def_Last_Linepack} \\
    & \sum_{d\in\mathcal{D}} \! \underline{M}_{a,d} x_{a,d} \leq m_{t,a} \leq \sum_{d\in\mathcal{D}} \! \overline{M}_{a,d} x_{a,d} 
    \quad \forall t,a
    \label{eqn:TP_Disc_Bounds_Linepack} 
\end{align} 
\end{subequations}

In~\eqref{eqn:TP_Disc_Linepack}, constraints~\eqref{eqn:TP_Disc_Flow_Linepack} limit the bidirectional gas inflow and outflow. Constraint~\eqref{eqn:TP_Disc_Def_Linepack} enforces mass conservation across consecutive time steps, thereby approximating pipeline linepack, while~\eqref{eqn:TP_Disc_Def_Last_Linepack} cyclically links the final and first time step, and~\eqref{eqn:TP_Disc_Bounds_Linepack} imposes lower and upper bounds on linepack.
The transport-linepack model constitutes a MILP with time-linking constraints.

In the following section, we solve the HEP problem and show how gas flow modeling choices affect model outcomes.

\section{Numerical Experiments}
\label{sec:Experiments}
This section presents numerical results for the HEP problem~\eqref{eqn:HEP}. We first describe the case study setup. Subsection~\ref{sec:Planning} then analyzes planning outcomes under different gas flow modeling choices across hydrogen demand levels, followed by validation under dynamic operation in~\ref{sec:Regret}.

The hydrogen network topology for the case study is based on a modified version of the GasLib 11-node gas network~\cite{Schmidt2017}, shown in Fig.~\ref{fig:Network}.
All pipelines have a length $L_a$ of \qty{60}{\kilo\meter}. Candidate pipeline diameters (\unit{\meter}) are $D_{a,d}=\{0.5,0.6,0.7,0.8,0.9,1.0,1.1,1.2,1.4\}$, with investment costs adopted from~\cite{Reuss2019} and annualized over a 40-year lifetime. The minimum and maximum system pressures are \qty{40}{\bar} and \qty{70}{\bar}, respectively, and the compressor ratio is $\Gamma_c=\qty{1.4}{}$.

Hydrogen production is represented by \qty{100}{MW} proton exchange membrane electrolyzers (PEMEL) directly coupled with renewable generation, with $\overline{P}^{E}_{t,e}$ representing the time-dependent available capacity of the electrolyzers due to renewable generation.
The maximum number of installable PEMEL units, $\overline{X}_{e}$, is limited to \num{800} wind-coupled units at node $i_1$ and \num{2000} solar-coupled units at node $i_8$ with associated annualized investment costs, $C^{\mathrm{Inv}}_e$, of \qty{3}{\mega\sieuro} per \num{100}-\unit{MW} plant~\cite{Reksten2022}. The electrolyzers convert electricity to hydrogen with an efficiency of $\eta_e= \qty{18.2}{\kg/\MW\hour}$. The electricity cost, $C^{\mathrm{E}}_{t,e}$, is \qty{2.5}{} for wind and \qty{1}{\sieuro/\mega\watt\hour} for solar generation. Finally, the cost of non-supplied hydrogen, $C^{\mathrm{ns}}$, is assumed as \qty{100}{\sieuro/\kg}.

Time series for $\overline{P}^{E}_{t,e}$ are calculated from normalized renewable generation from~\cite{REE2026}. Hydrogen demands, $D^{H}_{t,i}$, located at nodes $i_{7}$, $i_{10}$ and $i_{11}$, are derived from normalized industrial gas demand data from~\cite{NGT2026} and scaled to the test system.
Due to the computational complexity of the HEP problem, the test cases consider a single representative day with hourly resolution, i.e., $\Delta=\qty{1}{\hour}$, to represent annual system operation.

All models were solved in GAMS~52.5.0 on an Intel Core i9 workstation with 128~GB of RAM using an optimality gap of \qty{1e-6}{}. 
Depending on the computational complexity of the HEP problem arising from the gas flow formulation, different solvers were employed. SCIP~9.2.4 was used to solve expansion planning instances under the steady-state, transport, and transport-linepack models, as well as operational instances of the dynamic model (fixed investments), to global optimality. For expansion planning instances with the dynamic model, locally optimal solutions were obtained using DICOPT~2 with Gurobi~13.0.0 (MIP) and CONOPT~4.39 (NLP).

\definecolor{injection}{HTML}{2b8cbe}
\definecolor{withdrawal}{HTML}{a8ddb5}


\tikzset{
  circle node/.style={circle, draw, minimum size=2mm, inner sep=0pt, outer sep=0pt},
  triangle up/.style={regular polygon, regular polygon sides=3, draw, minimum size=3mm,
                        inner sep=0pt, outer sep=0pt, fill=injection, rotate=0 },
  triangle dn/.style={regular polygon, regular polygon sides=3, draw, minimum size=3mm,
                        inner sep=0pt, outer sep=0pt, fill=withdrawal, shape border rotate=180 },
  compressor/.style={
    circle,
    draw,
    minimum size=3mm,   
    inner sep=0pt,
    path picture={
      \draw
        ($(path picture bounding box.center)+(+1.5mm,+0.2mm)$) -- 
        ($(path picture bounding box.center)+(-0.8mm,+1.25mm)$);
      \draw
        ($(path picture bounding box.center)+(+1.5mm,-0.2mm)$) -- 
        ($(path picture bounding box.center)+(-0.8mm,-1.25mm)$);
    }
  },
  pipe/.style={line width=0.5pt, draw }
}

\newcommand{\networkscale}{1.0} 

\newcommand{\drawnetwork}[2]{%
\begin{scope}[scale=#2, transform shape]
  \begin{scope}[local bounding box=#1network]
    \node[triangle up, label=below:{$i_{1}$}                                 ] (1)  at (0,0) {}; 
    \node[circle node, label=below:{$i_{2}$},        right=1cm of 1          ] (2)  {};
    \node[compressor,                                right=0.2cm of 2        ] (C1) {}; 
    \node[circle node, label=below:{$i_{3}$},        right=0.2cm of C1       ] (3)  {};
    \node[circle node, label=right:{$i_{4}$},  above right=0.5cm and 1cm of 3] (4)  {};
    \node[circle node, label=right:{$i_{5}$},  below right=0.5cm and 1cm of 3] (5)  {};
    \node[circle node, label=below:{$i_{6}$},  above right=0.5cm and 1cm of 5] (6)  {};
    \node[triangle dn, label=right:{$i_{7}$},  above      =0.5cm         of 4] (7)  {};
    \node[triangle up, label=right:{$i_{8}$},  below      =0.5cm         of 5] (8)  {};
    \node[compressor,                                right=0.2cm of 6        ] (C2) {};    
    \node[circle node, label=below:{$i_{9}$},        right=0.2cm of C2       ] (9)  {};
    \node[triangle dn, label=below:{$i_{10}$}, above right=0.5cm and 1cm of 9] (10) {};
    \node[triangle dn, label=below:{$i_{11}$}, below right=0.5cm and 1cm of 9] (11) {};

    \pgfplotstablegetrowsof{\networkdata}
    \pgfmathtruncatemacro{\LastRow}{\pgfplotsretval-1}

    \foreach \r in {0,...,\LastRow}{%
        \pgfplotstablegetelem{\r}{start}\of\networkdata
        \edef\startnode{\pgfplotsretval}
        \pgfplotstablegetelem{\r}{end}\of\networkdata
        \edef\endnode{\pgfplotsretval}
        \pgfplotstablegetelem{\r}{decision}\of\networkdata
        \pgfmathtruncatemacro{\dec}{\pgfplotsretval}

        \ifnum\dec=1
          \draw[pipe, draw=black] (\startnode) -- (\endnode);
        \fi
    }
  \end{scope}
\end{scope}
}

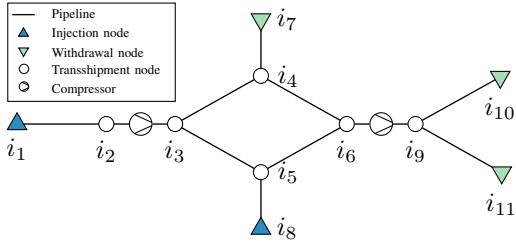
\begin{figure}[t]
\centering
\begin{tikzpicture}[nodes={align=center}]
\pgfplotstableread[col sep=comma]{figures/data/BaseNetwork.csv}\networkdata

\matrix[column sep=5mm, row sep=3mm] (mosaic) {
|[draw, thick, inner sep=1mm]| {\drawnetwork{A}{\networkscale}
    \node[font=\footnotesize, anchor=south] at (Anetwork.north) [yshift=0.0cm] {};} \\
};

\tikzset{
  triangle up legend/.style={regular polygon, regular polygon sides=3, draw, minimum size=3mm, inner sep=0pt, fill=injection},
  triangle dn legend/.style={regular polygon, regular polygon sides=3, draw, minimum size=3mm, inner sep=0pt, shape border rotate=180, fill=withdrawal},
  circle legend/.style={circle, draw, minimum size=2.5mm, inner sep=0pt},
  compressor legend/.style={
    circle,
    draw,
    minimum size=2.5mm,   
    inner sep=0pt,
    path picture={
      \draw
        ($(path picture bounding box.center)+(+1.5mm,+0.2mm)$) -- 
        ($(path picture bounding box.center)+(-0.8mm,+1.25mm)$);
      \draw
        ($(path picture bounding box.center)+(+1.5mm,-0.2mm)$) -- 
        ($(path picture bounding box.center)+(-0.8mm,-1.25mm)$);
    }
  }
}

\node[anchor=north west, draw, inner sep=0.5mm,
      xshift=2.5mm, yshift=-4mm] at (mosaic.north west)
{
\begin{tikzpicture}[scale=0.5, transform shape]

  \draw[pipe] (0,0) -- (0.6,0);
  \node[anchor=west] at (1,0) {Pipeline};

  \node[triangle up legend] (inj) at (0.24,-0.44) {};
  \node[anchor=west] at (1,-0.5) {Injection node};

  \node[triangle dn legend] (with) at (0.22,-0.88) {};
  \node[anchor=west] at (1,-1.0) {Withdrawal node};

  \node[circle legend] at (0.22,-1.35) {};
  \node[anchor=west] at (1,-1.5) {Transshipment node};

  \node[compressor legend] at (0.22,-1.82) {};
  \node[anchor=west] at (1,-2.0) {Compressor};

\end{tikzpicture}
};

\end{tikzpicture}
\caption{Hydrogen system based on GasLib-11 network topology.}
\label{fig:Network}
\end{figure}

\definecolor{ModelDynamic}{HTML}{1F77B4}
\definecolor{ModelSteady}{HTML}{FF7F0E}
\definecolor{ModelTransport}{HTML}{2CA02C}
\definecolor{ModelLinepack}{HTML}{D62728}

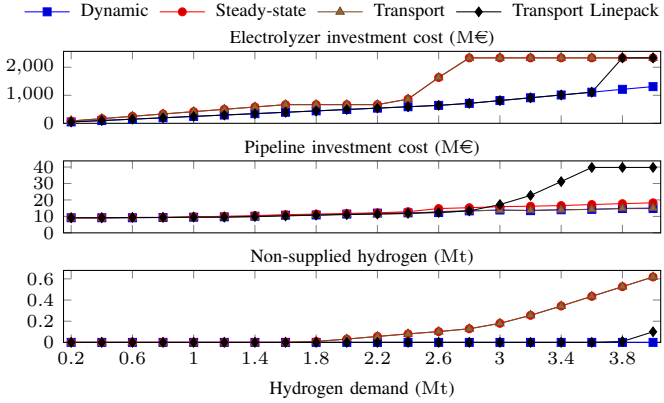
\begin{figure}[!t]
\centering
\pgfplotstableread[col sep=comma]{figures/data/Regret.csv}\costdata
\begin{tikzpicture}
\begin{groupplot}[
  group style={group size=1 by 3, horizontal sep=0cm, vertical sep=0.5cm},
  width=0.525\textwidth,
  height=0.14\textwidth,
  tick label style={font=\scriptsize},
  label style={font=\scriptsize},
  title style={font=\scriptsize, yshift=-0.8em},
  xmin=0.15, xmax=4.05,
  xtick={0.2,0.6,...,4},
  ymin=0,
  scaled y ticks=true,
  scaled x ticks=false,
  grid=none,
  major tick length=3pt,
  legend style={draw=none, font=\scriptsize, at={(0.47,1.28)}, anchor=south, legend columns=4, column sep=2pt},
  legend cell align={left},
]

\nextgroupplot[ylabel={}, title={Electrolyzer investment cost (\unit{\mega\sieuro})}, xticklabels={}]
\addplot+[mark=square*, mark size=1.5] table[x=Demand,y=Dynamic_PEMEL]{\costdata}; \addlegendentry{Dynamic}
\addplot+[mark=*, mark size=1.5] table[x=Demand,y=Steady_PEMEL]{\costdata}; \addlegendentry{Steady-state}
\addplot+[mark=triangle*, mark size=1.8] table[x=Demand,y=Transport_PEMEL]{\costdata}; \addlegendentry{Transport}
\addplot+[mark=diamond*, mark size=1.8] table[x=Demand,y=Linepack_PEMEL]{\costdata}; \addlegendentry{Transport Linepack}

\nextgroupplot[ylabel={}, title={Pipeline investment cost (\unit{\mega\sieuro})}, xticklabels={}]
\addplot+[mark=square*, mark size=1.5] table[x=Demand,y=Dynamic_Pipe]{\costdata}; 
\addplot+[mark=*, mark size=1.5] table[x=Demand,y=Steady_Pipe]{\costdata}; 
\addplot+[mark=triangle*, mark size=1.8] table[x=Demand,y=Transport_Pipe]{\costdata}; 
\addplot+[mark=diamond*, mark size=1.8] table[x=Demand,y=Linepack_Pipe]{\costdata}; 

\nextgroupplot[ylabel={}, title={Non-supplied hydrogen (\unit{\mega\tonne})}, xlabel={Hydrogen demand (\unit{\mega\tonne})}]
\addplot+[mark=square*, mark size=1.5] table[x=Demand,y=Dynamic_H2NS]{\costdata};
\addplot+[mark=*, mark size=1.5] table[x=Demand,y=Steady_H2NS]{\costdata};
\addplot+[mark=triangle*, mark size=1.8] table[x=Demand,y=Transport_H2NS]{\costdata};
\addplot+[mark=diamond*, mark size=1.8] table[x=Demand,y=Linepack_H2NS]{\costdata};

\end{groupplot}
\end{tikzpicture}
\caption{Electrolyzer and pipeline investment costs and non-supplied hydrogen under considered gas flow modeling choices across hydrogen demand levels.}
\label{fig:PlanningResults}
\end{figure}
\definecolor{injection}{HTML}{2b8cbe}
\definecolor{withdrawal}{HTML}{a8ddb5}

\definecolor{comp}{HTML}{ffffff}
\definecolor{500}{HTML}{fde725}
\definecolor{600}{HTML}{addc30}
\definecolor{700}{HTML}{5ec962}
\definecolor{800}{HTML}{28ae80}
\definecolor{900}{HTML}{21918c}
\definecolor{1000}{HTML}{2c728e}
\definecolor{1100}{HTML}{3b528b}
\definecolor{1200}{HTML}{472d7b}
\definecolor{1400}{HTML}{440154}

\tikzset{
  circle node/.style={circle, line width=0.2pt, draw, minimum size=2.2mm, inner sep=0pt, outer sep=0.2pt},
  triangle up/.style={regular polygon, regular polygon sides=3, line width=0.2pt, draw, minimum                                 size=3.5mm, inner sep=0pt, outer sep=0.2pt, fill=none, rotate=0 },
  triangle dn/.style={regular polygon, regular polygon sides=3, line width=0.2pt, draw, minimum                                 size=3.5mm, inner sep=0pt, outer sep=0.2pt, fill=none, shape border rotate=180 },
  compressor/.style={
    circle,
    line width=0.2pt,
    draw,
    minimum size=3.5mm,
    inner sep=0pt,
    path picture={
      \draw
        ($(path picture bounding box.center)+(+0.9mm,+0.1mm)$) --
        ($(path picture bounding box.center)+(-0.5mm,+0.75mm)$);
      \draw
        ($(path picture bounding box.center)+(+0.9mm,-0.1mm)$) --
        ($(path picture bounding box.center)+(-0.5mm,-0.75mm)$);
    }
  },
  pipe/.style={line width=0.8pt, draw}
}

\newcommand{\networkmosaicscale}{0.35}
\newcommand{\networkmosaicheaderone}{\qty{1}{\mega\tonne}}
\newcommand{\networkmosaicheadertwo}{\qty{2}{\mega\tonne}}
\newcommand{\networkmosaicheaderthree}{\qty{3}{\mega\tonne}}
\newcommand{\networkmosaicheaderfour}{\qty{4}{\mega\tonne}}
\newcommand{\networkmosaicrowheaderone}{Dynamic}
\newcommand{\networkmosaicrowheadertwo}{\shortstack{Steady-\\state}}
\newcommand{\networkmosaicrowheaderthree}{Transport}
\newcommand{\networkmosaicrowheaderfour}{\shortstack{Transport\\Linepack}}

\newcommand{\networkmosaiccolumnheader}[1]{%
  \ifcase#1\relax
    \networkmosaicheaderone
  \or
    \networkmosaicheadertwo
  \or
    \networkmosaicheaderthree
  \or
    \networkmosaicheaderfour
  \else
    Column \the\numexpr#1+1\relax
  \fi
}

\newcommand{\networkmosaicrowheader}[1]{%
  \ifcase#1\relax
    \networkmosaicrowheaderone
  \or
    \networkmosaicrowheadertwo
  \or
    \networkmosaicrowheaderthree
  \or
    \networkmosaicrowheaderfour
  \else
    Row \the\numexpr#1+1\relax
  \fi
}

\newcommand{\drawnetworkmosaic}[4]{%
\begin{scope}[scale=#2, transform shape]
  \begin{scope}[local bounding box=#1network]
    \node[triangle up] (#11) at (0,0) {};
    \node[circle node, right=1cm of #11] (#12) {};
    \node[compressor, right=0.2cm of #12] (#1C1) {};
    \node[circle node, right=0.2cm of #1C1] (#13) {};
    \node[circle node, above right=0.5cm and 1cm of #13] (#14) {};
    \node[circle node, below right=0.5cm and 1cm of #13] (#15) {};
    \node[circle node, above right=0.5cm and 1cm of #15] (#16) {};
    \node[triangle dn, above=0.5cm of #14] (#17) {};
    \node[triangle up, below=0.5cm of #15] (#18) {};
    \node[compressor, right=0.2cm of #16] (#1C2) {};
    \node[circle node, right=0.2cm of #1C2] (#19) {};
    \node[triangle dn, above right=0.5cm and 1cm of #19] (#110) {};
    \node[triangle dn, below right=0.5cm and 1cm of #19] (#111) {};

    \pgfplotstablegetrowsof{\networkdata}
    \pgfmathtruncatemacro{\LastRow}{\pgfplotsretval-1}
    \foreach \rowidx in {0,...,\LastRow}{%
      \pgfplotstablegetelem{\rowidx}{start}\of\networkdata
      \edef\startraw{\pgfplotsretval}
      \edef\startnode{#1\startraw}
      \pgfplotstablegetelem{\rowidx}{end}\of\networkdata
      \edef\endraw{\pgfplotsretval}
      \edef\endnode{#1\endraw}
      \pgfplotstablegetelem{\rowidx}{decision}\of\networkdata
      \pgfmathtruncatemacro{\dec}{\pgfplotsretval}

      \def\edgediameter{1}

      \pgfplotstablegetrowsof{\mosaicedgedata}
      \pgfmathtruncatemacro{\LastEdgeRow}{\pgfplotsretval-1}
      \foreach \edgerow in {0,...,\LastEdgeRow}{%
        \pgfplotstablegetelem{\edgerow}{row}\of\mosaicedgedata
        \pgfmathtruncatemacro{\csvrow}{\pgfplotsretval}
        \pgfplotstablegetelem{\edgerow}{col}\of\mosaicedgedata
        \pgfmathtruncatemacro{\csvcol}{\pgfplotsretval}
        \ifnum\csvrow=#3\relax
          \ifnum\csvcol=#4\relax
            \pgfplotstablegetelem{\edgerow}{start}\of\mosaicedgedata
            \edef\csvstart{\pgfplotsretval}
            \pgfplotstablegetelem{\edgerow}{end}\of\mosaicedgedata
            \edef\csvend{\pgfplotsretval}
            \ifnum\pdfstrcmp{\csvstart}{\startraw}=0\relax
              \ifnum\pdfstrcmp{\csvend}{\endraw}=0\relax
                \pgfplotstablegetelem{\edgerow}{diameter}\of\mosaicedgedata
                \xdef\edgediameter{\pgfplotsretval}
              \fi
            \fi
          \fi
        \fi
      }

      \pgfmathtruncatemacro{\diamclass}{\edgediameter}
      \def\edgecolor{black}
      \ifcase\diamclass\relax
        \def\edgecolor{comp}
      \or\def\edgecolor{black}
      \or\def\edgecolor{black}
      \or\def\edgecolor{500}
      \or\def\edgecolor{600}
      \or\def\edgecolor{700}
      \or\def\edgecolor{800}
      \or\def\edgecolor{900}
      \or\def\edgecolor{1000}
      \or\def\edgecolor{1100}
      \or\def\edgecolor{1200}
      \or\def\edgecolor{1400}
      \else
        \def\edgecolor{black}
      \fi
      \ifnum\dec=1
        \draw[pipe, draw=\edgecolor,
              preaction={draw=black, line width=1.2pt}] (\startnode) -- (\endnode);
      \fi
    }
  \end{scope}
\end{scope}
}

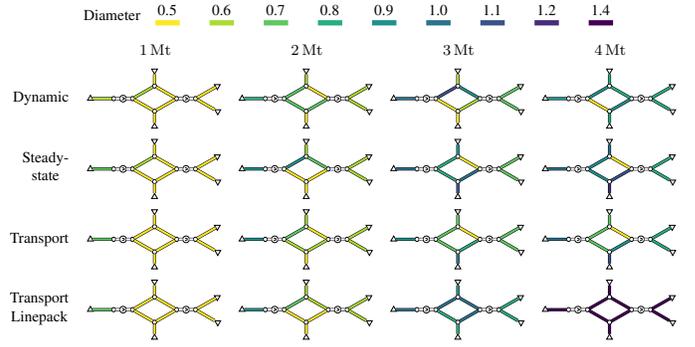
\begin{figure}[t]
\centering
\pgfplotstableread[col sep=comma]{figures/data/BaseNetwork.csv}\networkdata
\pgfplotstableread[col sep=comma]{figures/data/NetworkResults.csv}\mosaicedgedata
\begin{tikzpicture}[scale=0.72, transform shape]
\node[font=\footnotesize, anchor=base] at (0.4cm,1.45cm) {Diameter};
\foreach \idx/\clr/\lbl in {
  0/500/0.5,
  1/600/0.6,
  2/700/0.7,
  3/800/0.8,
  4/900/0.9,
  5/1000/1.0,
  6/1100/1.1,
  7/1200/1.2,
  8/1400/1.4
}{
  \draw[draw=\clr, line width=2pt] ({1.2cm + \idx*1cm},1.4cm) -- ({1.65cm + \idx*1cm},1.4cm);
  \node[font=\footnotesize, anchor=base] at ({1.42cm + \idx*1cm},1.55cm) {\lbl};
}

\foreach \col in {0,...,3}{
  \node[font=\footnotesize, anchor=base, inner sep=0pt, outer sep=0pt]
    at ({\col*2.8cm+1.2cm},0.8cm) {\strut\networkmosaiccolumnheader{\col}};
}
\foreach \row/\y in {0/0,1/1,2/2,3/3}{
  \node[font=\footnotesize, anchor=east, inner sep=0pt, outer sep=0pt]
    at (-0.4cm,{-\y*1.3cm+0.0cm}) {\strut\networkmosaicrowheader{\row}};
}

\foreach \row/\y in {0/0,1/1,2/2,3/3}{
  \foreach \col in {0,...,3}{
    \begin{scope}[shift={({\col*2.8cm},{-\y*1.3cm})}]
      \drawnetworkmosaic{N\row\col}{\networkmosaicscale}{\row}{\col}
    \end{scope}
  }
}
\end{tikzpicture}
\caption{Expanded pipeline diameters under considered gas flow modeling choices and hydrogen demand levels.}
\label{fig:NetworkResults}
\end{figure}

\subsection{Planning under Different Gas Flow Modeling Choices}
\label{sec:Planning}
We solve expansion planning instances of the HEP problem under different gas flow modeling choices. Fig.~\ref{fig:PlanningResults} reports electrolyzer and pipeline investment costs (\unit{\mega\sieuro}) and non-supplied hydrogen (\unit{\mega\tonne}) across demand levels, while Fig.~\ref{fig:NetworkResults} shows the corresponding pipeline expansions for selected instances.

For electrolyzer investments, the dynamic and transport-linepack models consistently yield lower costs than the steady-state and transport models. This reflects the operational flexibility provided by linepack, which enables expansion at locations with favorable hydrogen production conditions (i.e., renewable electricity availability and cost). The steady-state and transport models yield identical electrolyzer expansion.

Pipeline expansion varies across gas flow modeling choices. Pipeline investment costs under the steady-state model exceed those of the transport model across all demand levels, as expected due to its explicit representation of pressure–flow relationships. At high demand levels, the transport-linepack model favors larger pipeline diameters -- reaching the maximum expandable diameter due to its approximate representation -- an effect that is significantly less pronounced in the dynamic model for the considered demand levels (see Fig.~\ref{fig:NetworkResults}).

Importantly, from a demand of \qty{1.8}{\mega\tonne}, the steady-state and transport models begin to incur non-supplied hydrogen, whereas this occurs only beyond \qty{3.8}{\mega\tonne} in the transport-linepack model; the dynamic model maintains zero non-supplied hydrogen across all instances. These results clearly show the immense value of incorporating linepack flexibility in expansion planning models.

\subsection{Validation of Simplified Planning Decisions}
\label{sec:Regret}
We assess the quality of planning decisions obtained from the simplified models in Subsection~\ref{sec:Planning} when operated under the dynamic model using regret. To this end, let $J^{\mathrm{S}}(\cdot)$ and $J^{\mathrm{D}}(\cdot)$ denote the objective values of the simplified and dynamic models, respectively. Let $z^{\mathrm{op}}$ and $z^{\mathrm{inv}}$ denote operational and investment decisions, and let $\hat z^{\mathrm{inv}}$ be the optimal investments of a simplified model. 
The optimal objective values are defined as
\begin{align}
J^{\mathrm{S},*} &= \min_{z^{\mathrm{op}},z^{\mathrm{inv}}} J^{\mathrm{S}}(z^{\mathrm{op}}, z^{\mathrm{inv}}), \nonumber \\
J^{\mathrm{D},*}(\hat z^{\mathrm{inv}}) &= \min_{z^{\mathrm{op}}} J^{\mathrm{D}}(z^{\mathrm{op}}, \hat z^{\mathrm{inv}}), \nonumber \\
J^{\mathrm{D},*}
&= \min_{z^{\mathrm{op}},z^{\mathrm{inv}}} J^{\mathrm{D}}(z^{\mathrm{op}}, z^{\mathrm{inv}}),
\end{align}
and the relative regret is defined as
\begin{equation}
\mathrm{Regret} = 
\frac{J^{\mathrm{D},*}(\hat z^{\mathrm{inv}})-J^{\mathrm{D},*}}{J^{\mathrm{D},*}}.
\label{eq:regret-main}
\end{equation}

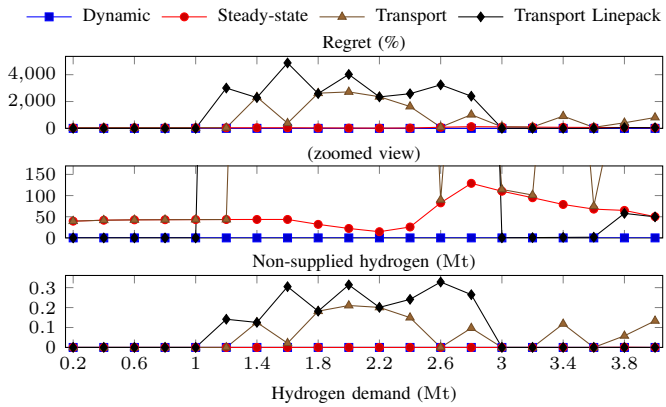
\begin{figure}[!t]
\centering
\pgfplotstableread[col sep=comma]{figures/data/Regret.csv}\costdata
\begin{tikzpicture}
\begin{groupplot}[
  group style={group size=1 by 3, horizontal sep=0cm, vertical sep=0.5cm},
  width=0.525\textwidth,
  height=0.14\textwidth,
  tick label style={font=\scriptsize},
  label style={font=\scriptsize},
  title style={font=\scriptsize, yshift=-0.8em},
  xmin=0.15, xmax=4.05,
  xtick={0.2,0.6,...,4},
  ymin=0,
  scaled y ticks=true,
  scaled x ticks=false,
  grid=none,
  major tick length=3pt,
  legend style={draw=none, font=\scriptsize, at={(0.47,1.28)}, anchor=south, legend columns=4, column sep=2pt},
  legend cell align={left},
]

\nextgroupplot[ylabel={}, title={Regret (\%)}, xticklabels={}]
\addplot+[mark=square*, mark size=1.5] table[x=Demand,y=Dynamic_Regret]{\costdata}; \addlegendentry{Dynamic}
\addplot+[mark=*, mark size=1.5] table[x=Demand,y=Steady_Regret]{\costdata}; \addlegendentry{Steady-state}
\addplot+[mark=triangle*, mark size=1.8] table[x=Demand,y=Transport_Regret]{\costdata}; \addlegendentry{Transport}
\addplot+[mark=diamond*, mark size=1.8] table[x=Demand,y=Linepack_Regret]{\costdata}; \addlegendentry{Transport Linepack}

\nextgroupplot[ylabel={}, title={(zoomed view)}, ymax=170, xticklabels={}]
\addplot+[mark=square*, mark size=1.5] table[x=Demand,y=Dynamic_Regret]{\costdata};
\addplot+[mark=*, mark size=1.5] table[x=Demand,y=Steady_Regret]{\costdata}; 
\addplot+[mark=triangle*, mark size=1.8] table[x=Demand,y=Transport_Regret]{\costdata};
\addplot+[mark=diamond*, mark size=1.8] table[x=Demand,y=Linepack_Regret]{\costdata}; 

\nextgroupplot[ylabel={}, title={Non-supplied hydrogen (\unit{\mega\tonne})}, xlabel={Hydrogen demand (\unit{\mega\tonne})}]
\addplot+[mark=square*, mark size=1.5] table[x=Demand,y=Dynamic_RegretH2NS]{\costdata};
\addplot+[mark=*, mark size=1.5] table[x=Demand,y=Steady_RegretH2NS]{\costdata};
\addplot+[mark=triangle*, mark size=1.8] table[x=Demand,y=Transport_RegretH2NS]{\costdata};
\addplot+[mark=diamond*, mark size=1.8] table[x=Demand,y=Linepack_RegretH2NS]{\costdata};

\end{groupplot}
\end{tikzpicture}
\caption{Regret from planning under simplified gas flow models under dynamic operation (top), with a zoomed view (middle), and non-supplied hydrogen (bottom), across hydrogen demand levels.}
\label{fig:Regret}
\end{figure}

Fig.~\ref{fig:Regret} shows regret (\qty{}{\%}) and non-supplied hydrogen (\qty{}{\mega\tonne}) resulting from planning under simplified gas flow models and subsequent operation under the more realistic dynamic model. The results demonstrate that approximating pressure-flow relationships in the transport and transport-linepack models constitutes a clear over-simplification of physical gas flow, leading to suboptimal system expansion that cannot be compensated by linepack flexibility in the dynamic model. Notably, the transport-linepack model overestimates linepack flexibility, yielding instances with near-zero regret -- particularly at low demand levels -- while also exhibiting the maximum regret (\qty{4875}{\%}) across all instances. Importantly, planning under these simplified models does not yield robust expansion plans, as instances without non-supplied hydrogen in planning may incur it in operation, and vice versa.
Planning under the steady-state model does not achieve zero regret in any instance. However, this regret primarily stems from suboptimal investment and operational decisions rather than non-supplied hydrogen (except for the \qty{3.8}{\mega\tonne} instance, where \qty{1}{\kilo\tonne} of demand is unmet). Overall, the steady-state model yields more consistent regret across demand levels, with a maximum of \qty{129}{\%}, indicating more robust expansion plans. Still, its average regret of \qty{56}{\%} implies significant untapped cost-reduction potential in system planning.

\definecolor{ModelDynamic}{HTML}{1F77B4}
\definecolor{ModelSteady}{HTML}{FF7F0E}
\definecolor{ModelTransport}{HTML}{2CA02C}
\definecolor{ModelLinepack}{HTML}{D62728}

\section{Conclusions}
\label{sec:Conclusions}
This paper analyzes the implications of simplified pipeline gas flow models for integrated energy system planning, as commonly used in the power systems community. We formulate a power–hydrogen expansion planning model with gas flows represented by steady-state, transport, and transport models with linepack, and evaluate the resulting planning decisions under a more realistic dynamic model.

Our results show that planning under the transport and transport-linepack models can yield expansion plans with significant non-supplied hydrogen due to the over-simplification of gas flow physics and, importantly, lack of robustness across demand levels. In our case study, the steady-state model avoids non-supplied hydrogen in all but one instance. However, it still incurs substantial regret due to suboptimal investment and operation, averaging \qty{56}{\%} relative to the dynamic model. While computationally demanding, the steady-state model offers a favorable trade-off between accuracy and tractability among the considered approaches.

Overall, this work provides insights for the power systems modeling community to reconsider the default use of over-simplified transport and transport-linepack models and instead adopt more realistic steady-state formulations, which are state-of-the-art in the gas systems community for network planning. Although highly promising for integrated energy system planning, the dynamic model remains computationally highly demanding even for relatively small instances. Future work should therefore focus on developing computationally efficient solution algorithms for the dynamic model to accurately capture gas dynamics in expansion planning problems.

\bibliographystyle{bib/IEEEtran}
\bibliography{bib/main}

@article{Schmidt2017,
author = {Schmidt, Martin and A{\ss}mann, Denis and Burlacu, Robert and Humpola, Jesco and Joormann, Imke and Kanelakis, Nikolaos and Koch, Thorsten and Oucherif, Djamal and Pfetsch, Marc E. and Schewe, Lars and Schwarz, Robert and Sirvent, Mathias},
journal = {Data},
number = {4},
pages = {1--18},
title = {{GasLib—A library of gas network instances}},
volume = {2},
year = {2017}
}

@article{Raheli2025,
author = {Raheli, Enrica and Werner, Yannick and Kazempour, Jalal},
journal = {IEEE Trans. Power Syst.},
month = {May},
number = {3},
pages = {2130--2142},
title = {{Flexibility of Integrated Power and Gas Systems: Modeling and Solution Choices Matter}},
volume = {40},
year = {2025}
}

@phdthesis{Lenz2021,
author = {Lenz, Ralf},
school = {Technische Universit{\"{a}}t Berlin},
title = {{Optimization of Stationary Expansion Planning and Transient Network Control by Mixed-Integer Nonlinear Programming}},
type = {Dissertation},
year = {2021}
}

@book{Koch2015,
address = {Philadelphia, PA},
author = {Koch, Thorsten and Hiller, Benjamin and {E. Pfetsch}, Marc and Schewe, Lars},
booktitle = {Evaluating Gas Network Capacities},
doi = {10.1137/1.9781611973693},
publisher = {Society for Industrial and Applied Mathematics},
title = {{Evaluating Gas Network Capacities}},
year = {2015}
}

@article{Reuss2019,
author = {Reu{\ss}, Markus and Welder, Lara and Th{\"{u}}rauf, Johannes and Lin{\ss}en, Jochen and Grube, Thomas and Schewe, Lars and Schmidt, Martin and Stolten, Detlef and Robinius, Martin},
journal = {Int. J. Hydrogen Energy},
number = {60},
pages = {32136--32150},
title = {{Modeling hydrogen networks for future energy systems: A comparison of linear and nonlinear approaches}},
volume = {44},
year = {2019}
}

@article{Neumann2023,
    author = {Fabian Neumann and Elisabeth Zeyen and Marta Victoria and Tom Brown},
    title = {The potential role of a hydrogen network in Europe},
    journal = {Joule},
    volume = {7},
    pages = {1--25},
    year = {2023}
}

@article{Brown2018,
  author        = {Brown, Tom and Hörsch, Jonas and Schlachtberger, David},
  title         = {Py{PSA}: {P}ython for {P}ower {S}ystem {A}nalysis},
  journal       = {J. Open Res. Softw.},
  year          = {2018},
  volume        = {6},
  number        = {1},
  pages         = {4},
}

@techreport{EHB2023,
author = {{European Hydrogen Backbone Initiative}},
title = {{EHB initiative to provide insights on infrastructure development by 2030}},
year = {2023}
}

@misc{NGT2026,
author = {{National Gas Transmission}},
title = {{Industrial Demand}},
url = {https://data.nationalgas.com/reports/customisable-downloads},
urldate = {2026-02-26},
year = {2026}
}

@misc{REE2026,
author = {{Red El{\'{e}}ctrica de Espa{\~{n}}a}},
title = {{CO2-free Generation}},
url = {https://www.esios.ree.es/en/analysis},
urldate = {2026-02-26},
year = {2026}
}

@misc{EUEnergySystemIntegration2020,
author = {{European Commission}},
booktitle = {European Commission},
title = {{Powering a climate-neutral economy: An EU Strategy for Energy System Integration}},
url = {https://eur-lex.europa.eu/legal-content/EN/ALL/?uri=COM:2020:299:FIN},
year = {2020}
}

@article{Baufume2013,
author = {Baufum{\'{e}}, Sylvestre and Gr{\"{u}}ger, Fabian and Grube, Thomas and Krieg, Dennis and Linssen, Jochen and Weber, Michael and Hake, J{\"{u}}rgen Friedrich and Stolten, Detlef},
journal = {Int. J. Hydrogen Energy},
number = {10},
pages = {3813--3829},
title = {{GIS-based scenario calculations for a nationwide German hydrogen pipeline infrastructure}},
volume = {38},
year = {2013}
}

@article{Andre2013,
author = {Andr{\'{e}}, Jean and Auray, St{\'{e}}phane and Brac, Jean and {De Wolf}, Daniel and Maisonnier, Guy and Ould-Sidi, Mohamed Mahmoud and Simonnet, Antoine},
journal = {Eur. J. Oper. Res.},
number = {1},
pages = {239--251},
title = {{Design and dimensioning of hydrogen transmission pipeline networks}},
volume = {229},
year = {2013}
}

@article{Zeng2017,
author = {Zeng, Qing and Zhang, Baohua and Fang, Jiakun and Chen, Zhe},
journal = {Appl. Energy},
month = {Aug},
pages = {192--203},
title = {{A bi-level programming for multistage co-expansion planning of the integrated gas and electricity system}},
volume = {200},
year = {2017}
}

@inproceedings{Borraz-Sanchez2016a,
address = {Hawaii, USA},
author = {Sanchez, Conrado Borraz and Bent, Russell and Backhaus, Scott and Blumsack, Seth and Hijazi, Hassan and van Hentenryck, Pascal},
booktitle = {2016 49th Hawaii  Int. Conf. System Sciences (HICSS)},
title = {{Convex Optimization for Joint Expansion Planning of Natural Gas and Power Systems}},
year = {2016}
}

@article{Borraz-Sanchez2016,
author = {Borraz-S{\'{a}}nchez, Conrado and Bent, Russell and Backhaus, Scott and Hijazi, Hassan and {Van Hentenryck}, Pascal},
journal = {INFORMS J. Comput.},
number = {4},
pages = {645--656},
title = {{Convex relaxations for gas expansion planning}},
volume = {28},
year = {2016}
}

@article{Zhao2018,
author = {Zhao, Bining and Conejo, Antonio J. and Sioshansi, Ramteen},
journal = {IEEE Trans. Power Syst.},
month = {May},
number = {3},
pages = {3064--3075},
title = {{Coordinated Expansion Planning of Natural Gas and Electric Power Systems}},
volume = {33},
year = {2018}
}

@article{Reksten2022,
author = {Reksten, Anita H. and Thomassen, Magnus S. and M{\o}ller-Holst, Steffen and Sundseth, Kyrre},
journal = {Int. J. Hydrogen Energy},
number = {90},
pages = {38106--38113},
title = {{Projecting the future cost of PEM and alkaline water electrolysers; a CAPEX model including electrolyser plant size and technology development}},
volume = {47},
year = {2022}
}

@article{Bodal2024,
author = {B{\o}dal, Espen Flo and Holm, Sigmund Eggen and Subramanian, Avinash and Durakovic, Goran and Pinel, Dimitri and Hellemo, Lars and Ortiz, Miguel Mu{\~{n}}oz and Knudsen, Brage Rugstad and Straus, Julian},
journal = {Appl. Energy},
number = {May 2023},
pages = {122484},
title = {{Hydrogen for harvesting the potential of offshore wind: A North Sea case study}},
volume = {357},
year = {2024}
}

@article{He2021,
author = {He, Guannan and Mallapragada, Dharik S. and Bose, Abhishek and Heuberger, Clara F. and Gencer, Emre},
journal = {IEEE Trans. Sustainable Enery},
month = {Jul},
number = {3},
pages = {1730--1740},
title = {{Hydrogen Supply Chain Planning With Flexible Transmission and Storage Scheduling}},
volume = {12},
year = {2021}
}

@misc{EURFNBO2023,
  author        = {{European Commission}},
  title         = {{Commission Delegated Regulation (EU) 2023/1184 of 10 February 2023}},
  year          = {2023},
  howpublished  = {\url{http://data.europa.eu/eli/reg_del/2023/1184/oj}}
}

@article{Wogrin2022,
  author        = {Sonja Wogrin and Diego Alejandro Tejada-Arango and Robert Gaugl and Thomas Klatzer and Udo Bachhiesl},
  title         = {{LEGO}: {T}he open-source {L}ow-carbon {E}xpansion {G}eneration {O}ptimization model},
  journal       = {SoftwareX},
  year          = {2022},
  volume        = {19},
  pages         = {101141}, 
}

\end{document}